\documentclass[twocolumn,english,showpacs]{revtex4}
\usepackage{times}
\usepackage[OT1]{fontenc}
\usepackage[latin1]{inputenc}
\usepackage{amsmath}
\usepackage{graphicx}
\usepackage{amssymb}

\makeatletter
\usepackage{babel}
\makeatother
\begin{document}

\title{Adiabatic loading of bosons into optical lattices}

\author{P. B. Blakie and J. V. Porto}

\affiliation{Physics Laboratory, National Institute of Standards and Technology,
Gaithersburg, Maryland 20899, USA}

\date{\today}

\begin{abstract}
The entropy-temperature curves are calculated for non-interacting
bosons in a 3D optical lattice and a 2D lattice with transverse harmonic
confinement for ranges of depths and filling factors relevant to current
experiments. We demonstrate regimes where the atomic sample can be
significantly heated or cooled by adiabatically changing the lattice
depth. We indicate the critical points for condensation in the presence
of a lattice and show that the system can be reversibly condensed
by changing the lattice depth. We discuss the effects of interactions
on our results and consider non-adiabatic processes.
\end{abstract}

\pacs{03.75.Fi, 32.80.Pj, 05.30.-d}

\maketitle

\section{Introduction}

Neutral bosonic atoms in optical lattices have been used to demonstrate
quantum matter-wave engineering \cite{Orzel2001a,Greiner2002b}, the
Mott-insulator quantum-phase transition \cite{Greiner2002a}, and
explore quantum entanglement \cite{Mandel2003a}. The many favorable
attributes of optical lattices, such as the low noise level and high
degree of experimental control, make them an ideal system for implementing
quantum logic \cite{Calarco2000a,Deutsch2000a}. A central part of
many of these proposals is the use of the Mott-insulator transition
to prepare the system into fiducial state with precisely one atom
per site, and negligible quantum or thermal fluctuations.

The usual path for preparing a sample of quantum degenerate bosons
in an optical lattice consists of first forming a cold Bose-Einstein
condensate in a weak magnetic trap, to which a 1, 2 or 3D lattice
potential is adiabatically applied by slowly ramping up the light
field intensity. Ideally this process will transfer a condensate at
zero temperature into its many-body ground state in the lattice. Of
course condensates cannot be produced at $T=0$K, yet to date the
role of temperature has received little attention even though it may
play a crucial role in the many-body properties of these systems.
Indeed, due to the massive energy spectrum changes the system undergoes
in lattice loading, the initial and final temperatures are not trivially
related. It is also of interest to understand how $T_{c}$ changes
in the lattice to assess the effect the periodic potential has on
condensation. Relevant to these considerations, experiments reported
in \cite{Burger2002a} examined evaporative cooling of atoms in a
combined magnetic trap and 1D optical lattice, and showed a significant
decrease in the critical temperature for a relatively shallow lattice
depth. 

Using entropy comparison Olshanni and Weiss \cite{Olshanii2002a}
have considered how a thermalized system of bosons in an optical lattice
would be transformed through adiabatic unloading into simple traps,
with a view to producing a condensate optically. Their approach takes
into account spatially inhomogeneous potentials superimposed upon
the lattice, but they assume that each lattice site is occupied by
no more that one boson and that the tunneling rate between sites is
zero --- strictly valid only for infinitely deep lattices.

In this paper we begin by considering the thermodynamic properties
of an ideal gas of bosons in a 3D cubic lattice. Working with the
grand canonical ensemble we use the exact single particle eigenstates
of the lattice to determine the entropy-temperature curves for the
system for various lattice depths and filling factors. We use these
curves for analyzing the effect of the loading process on the temperature
of the system, and show that sufficiently cold atomic samples can
be significantly cooled through loading into a lattice. By analyzing
the nature of the energy spectrum we explain this counter-intuitive
notion that adiabatic compression of a system can lead to cooling. 

Interactions between particles are not accounted for in our calculations
of the thermodynamic properties, however by considering the Bogoliubov
excitation spectrum in the lattice we discuss the modifications interactions
should introduce to our results. We also address how robust our predictions
are to non-adiabatic effects in the loading procedure. 

In the last part of this paper we investigate the thermodynamic properties
of bosonic atoms in a two-dimensional lattice. An important experimental
consideration in this case is that the intensity envelope of the lasers
used to make the optical lattice gives rise to an additional slowly-varying
potential perpendicular to the plane of the lattice sites. For the
case of a red detuned lattice this potential is confining and approximately
harmonic in the region where the atoms are trapped. With increasing
laser intensity the lattice (ground band) degrees of freedom become
more degenerate, whereas the energy spacing of harmonic degrees of
freedom increases due to the strengthened transverse confinement.
Competition between these two effects considerably modifies the nature
of the heating or cooling that occurs during lattice loading. We model
the thermodynamic properties of this system and present numerical
calculations for the entropy-temperature curves for parameter regimes
relevant to current experiments.

\section{Formalism - 3D Lattice}

\subsection{Single Particle Eigenstates}

We consider a cubic 3D optical lattice made from 3 independent (i.e.
non-interfering) sets of counter-propagating laser fields of wavelength
$\lambda$, giving rise to a potential of the form\begin{equation}
V_{{\rm {Latt}}}(\mathbf{r})=\frac{V}{2}[\cos(2kx)+\cos(2ky)+\cos(2kz)],\label{eq:LattPot}\end{equation}
where $k=2\pi/\lambda$ is the single photon wavevector, and $V$
is the lattice depth. We take the lattice to be of finite extent with
a total of $N_{s}$ sites, consisting of an equal number of sites
along each of the spatial directions with periodic boundary conditions.
The single particle energies \textbf{$\epsilon_{\mathbf{q}}$} are
determined by solving the Schr\"odinger equation

\begin{equation}
\epsilon_{\mathbf{q}}\psi_{\mathbf{q}}(\mathbf{r})=\frac{\mathbf{p^{2}}}{2m}\psi_{\mathbf{q}}(\mathbf{r})+V_{{\rm {Latt}}}(\mathbf{r})\psi_{\mathbf{q}}(\mathbf{r}),\label{eq:BlochState}\end{equation}
 for the Bloch states, $\psi_{\mathbf{q}}(\mathbf{r}),$ of the lattice.
For notational simplicity we choose to work in the extended zone scheme
where $\mathbf{q}$ specifies both the quasimomentum and band index
of the state under consideration. By using the single photon recoil
energy, $E_{R}=\hbar^{2}k^{2}/2m$, as our unit of energy, the energy
states of the system are completely specified by the lattice depth
$V$ and the number of lattice sites $N_{s}$ (i.e. in recoil units
$\epsilon_{\mathbf{q}}$ is independent of $k$). 

It is useful to review the tight-binding description of the ground
band. Valid for moderately deep lattices, the tight-binding model
only accounts for nearest neighbor tunneling, and leads to the analytic
dispersion relation \begin{equation}
\epsilon_{\mathbf{q}}^{{\rm TB}}=\frac{\hbar^{2}}{m^{*}a^{2}}\left(3-\sum_{j=x,y,z}\cos(q_{j}a)\right),\label{eq:ETB}\end{equation}
where $m^{*}=\left(\hbar^{-2}\nabla_{\mathbf{q}}^{2}\epsilon_{\mathbf{q}}\Big|_{\mathbf{q=}0}\right)^{-1}$
is the effective mass at $\mathbf{q=0}$, and $a=\lambda/2$ is the
lattice site spacing.

\subsection{Equilibrium Properties}

Our interest lies in understanding the process of adiabatically loading
system of $N_{p}$ bosons into a lattice. (The requirements for adiabaticity
in this system are not well understood, though the timescales for
adiabatically changing the lattice depth are expected to become long
in deep lattices, which we discuss further in Sec. \ref{sub:Adiabaticity}).
Under the assumption of adiabaticity the entropy remains constant
throughout this process and the most useful information can be obtained
from knowing how the entropy depends on the other parameters of the
system. In the thermodynamic limit, where $N_{s}\to\infty$ and $N_{p}\to\infty$
while the filling factor $n\equiv N_{p}/N_{s}$ remains constant,
the entropy per particle is completely specified by the intensive
parameters $T,V,n$. The calculations we present in this paper are
for finite size systems, that are sufficiently large to approximate
the thermodynamic limit. 

To determine the entropy, the single particle spectrum $\{\epsilon_{\mathbf{q}}\}$
of the lattice is calculated for given values of $N_{s}$ and $V$.
We then determine the thermodynamic properties of the lattice with
$N_{p}$ bosons in the grand canonical ensemble, for which we calculate
the partition function $\mathcal{Z}$\begin{equation}
\log\mathcal{Z}=-\sum_{\mathbf{q}}\log\left(1-e^{-\beta(\epsilon_{\mathbf{q}}-\mu)}\right),\label{eq:GrandPot}\end{equation}
where $\mu$ is found by ensuring particle conservation. The entropy
of the system can then be expressed as\begin{equation}
S=k_{B}\left(\log\mathcal{Z}+\beta E-\mu\beta N_{p}\right),\label{eq:Entropy}\end{equation}
where $\beta=1/k_{B}T$, and $E=-\partial\ln\mathcal{Z}/\partial\beta$
is the mean energy.

\subsubsection{Density of states }

\begin{figure}
\includegraphics[%
  width=3.5in]{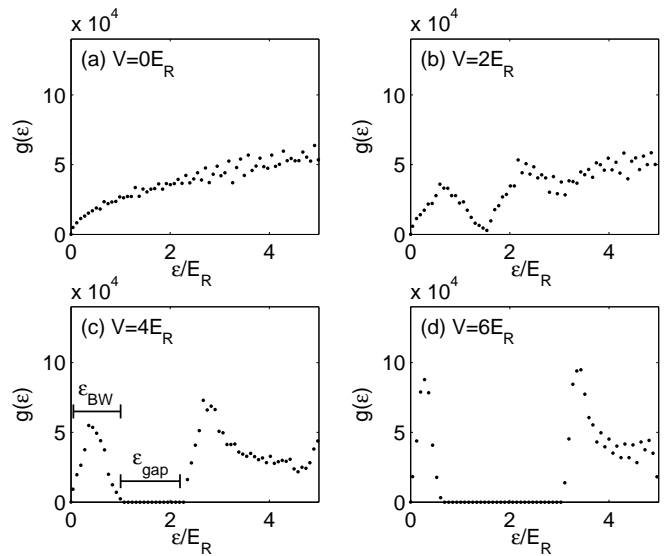}

\caption{\label{FIG:DOS} Density of states for a $N_{s}\approx3\times10^{4}$
site cubic lattice at various depths. For a depth of approximately
$V=2E_{R}$ a gap develops in the density of states. In (c) we illustrate
the energy gap, $\epsilon_{{\rm gap}}$, and ground bandwidth, $\epsilon_{{\rm BW}}$.
Points are determined by numerically averaging the exact spectrum
over a small energy range.}
\end{figure}

How the thermodynamic properties of a system of bosons change as they
are adiabatically loaded into a lattice intimately reflects how the
lattice modifies the microscopic energy spectrum. In this regard the
density of states function affords considerable insight into the behaviour
of the system. In Fig. \ref{FIG:DOS} we illustrate how the distribution
of available single particle states changes for various lattice depths.

In general we note that the lattice leads to a substantial change
in the the density of states, $g(\epsilon)$, for the system. In the
absence of the lattice (Fig. \ref{FIG:DOS}(a)) the density of states
is that for free particles in a box, and is proportional to $\sqrt{\epsilon}$.
The smoothness of $g(\epsilon)$ is disrupted by the presence of a
lattice, which causes flat regions in the energy bands giving rise
to peaked features in the density of states, known as \emph{van Hove
singularities} \cite{Mermin1976}. The van Hove singularities in the
first and second energy bands are clearly visible in Figs. \ref{FIG:DOS}(b)-(d).
For sufficiently deep lattices an energy gap, $\epsilon_{{\rm gap}}$,
will separate the ground and first excited bands. For the cubic lattice
we consider here, a finite gap appears at a lattice depth of  $V\approx2E_{R}$
\footnote{The delay in appearance of the excitation spectrum gap until $V\approx2E_{R}$
is a property of the 3D band structure. In a 1D lattice a gap is present for
all depths $V>0$.%
} (see Fig. \ref{FIG:DOS}(b)), and beyond this depth the gap increases
with lattice depth (see Figs. \ref{FIG:DOS}(c)-(d)). In forming the
gap higher energy bands are shifted up-wards in energy, and the ground
band becomes compressed --- a feature characteristic of the reduced
tunneling between lattice sites. We refer to the energy range over
which the ground band extends as the (ground) band width, $\epsilon_{{\rm BW}}$.
This quantity decreases in magnitude exponentially with $V$ (see
Figs.\ref{FIG:DOS}(b)-(c)), causing the ground band to have an extremely
high density of states for deep lattices; generally in this limit
quantum many-body effects will significantly modify the energy spectra
of the lowest band from the non-interacting states we use here. We
will discuss the effects of interactions in Sec. \ref{sub:Interaction-effects}.

\section{Results}

\begin{figure}
\includegraphics[%
  width=3.4in]{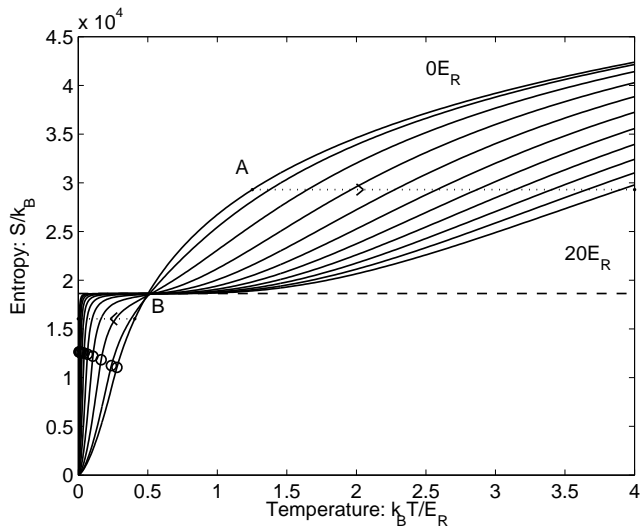}

\caption{\label{FIG:ST1} Entropy versus temperature curves for a $N_{s}\approx3\times10^{4}$
site cubic lattice, with filling factor $n=0.25$ at various depths
$V=0$ to $20E_{R}$ (with a spacing of $2E_{R}$ between each curve).
The entropy plateau $S_{0}$ is shown as a dashed line and the condensation
point is marked on each curve as a circle. Dotted line marked $A$
shows a path along which adiabatic loading into the lattice causes
the temperature to increase. Dotted line marked $B$ shows a path
along which adiabatic loading into the lattice causes the temperature
to decrease. }
\end{figure}

\begin{figure}
\includegraphics[%
  width=3.4in]{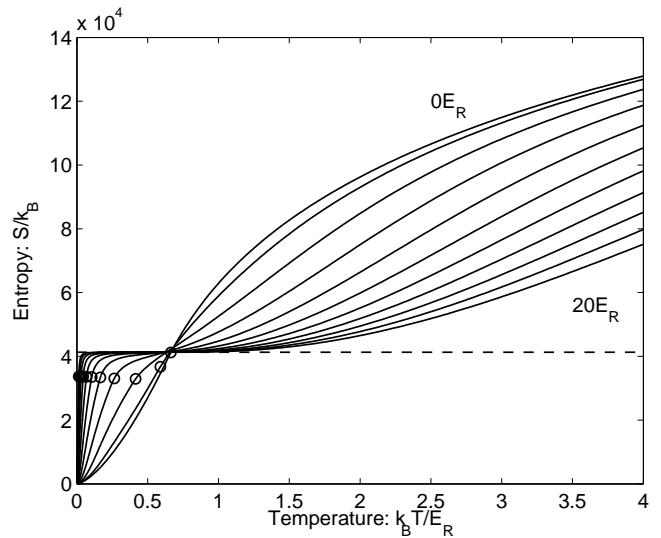}

\caption{\label{FIG:ST2} Entropy versus temperature curves for a $N_{s}\approx3\times10^{4}$
site cubic lattice, with filling factor $n=1$ at various depths$V=0$
to $20E_{R}$ (with a spacing of $2E_{R}$ between each curve). The
entropy plateau $S_{0}$ is shown as a dashed line and the condensation
point is marked on each curve as a circle.}
\end{figure}

\begin{figure}
\includegraphics[%
  width=3.4in]{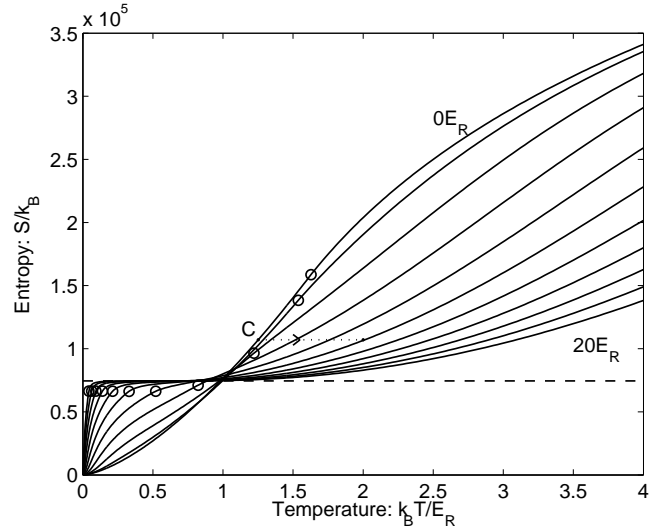}

\caption{\label{FIG:ST3} Entropy versus temperature curves for a $N_{s}\approx3\times10^{4}$
site cubic lattice, with filling factor $n=4$ at various depths $V=0$
to $20E_{R}$ (with a spacing of $2E_{R}$ between each curve). The
entropy plateau $S_{0}$ is shown as a dashed line and the condensation
point is marked on each curve as a circle. Dotted line marked $C$
shows a path along which an initially Bose-condensed system uncondenses
as it is loaded into the lattice at $V\approx4E_{R}$.}
\end{figure}

In Figs. \ref{FIG:ST1}-\ref{FIG:ST3} we show entropy-temperature
curves for various lattice depths and filling factors $n$. These
curves have been calculated for a lattice with $31$ lattice sites
along each spatial dimension, i.e. $N_{s}\approx3\times10^{4}$. The
condensation temperature is defined as that at which 1\% of all particle
occupy the ground state, and is indicated in Figs. \ref{FIG:ST1}-\ref{FIG:ST3}.
We note that being a finite system the transition is not discontinuous,
however the onset of condensation is rapid and changing the requirement
to 5\% makes little observable difference in the critical point locations.

An important common feature to these curves is the distinct separation
of regions where adiabatic loading causes the temperature of the sample
to increase or decrease, which we will refer to as the regions of
heating and cooling respectively. These two regions are separated
by a common point that all curves approximately pass through, and
we denote by its coordinates as $\{ T_{0},S_{0}\}$. The reason for
the existence of this point will be discussed below. For cases considered
in Figs. \ref{FIG:ST1}-\ref{FIG:ST3} $T_{0}$ is in the range $0.5-1E_{R}$,
and to clearly indicate the vertical separation of the heating and
cooling regions, we have marked $S_{0}$ as a horizontal dashed line. \textbf{}

We now explicitly demonstrate the temperature changes that occur during
adiabatic loading using two possible adiabatic processes labeled $A$
and $B$, and marked as dotted lines in Fig. \ref{FIG:ST1}. Process
$A$ begins with a gas of free particles in an initial state with
$S>S_{0},$ $T>T_{0}$. As the gas is loaded into the lattice the
process line indicates that the temperature increases rapidly with
the lattice depth. Conversely process $B$ begins with a gas of free
particles in an initial state with $S<S_{0},$ $T<T_{0},$ and the
lattice loading causes a rapid decrease in temperature. This behavior
can be qualitatively understood in terms of the modifications the
lattice makes to the energy states of the system. As is apparent in
Fig. \ref{FIG:DOS}, the ground band flattens for increasing lattice
depth causing the density of states to be more densely compressed
at lower energies. Thus in the lattice all these states can be occupied
at a much lower temperature than for the free particle case. As we
discuss below, $S_{0}$ is the maximum entropy available from only
accessing states of the lowest band, and so for $S>S_{0}$ higher
bands are important. As the lattice depth and hence $\epsilon_{{\rm gap}}$
increases, the temperature must increase for these excited states
to remain accessible.

\subsection{Entropy plateau\label{sub:Entropy-Plateau}}

In Figs. \ref{FIG:ST1}-\ref{FIG:ST3} a horizontal plateau (at the
level marked by the dashed lines) is common to the entropy-temperature
curves for larger lattice depths ($V\gtrsim8E_{R}$). This occurs
because for these lattices $\epsilon_{{\rm BW}}\ll\epsilon_{{\rm gap}}$,
and there is a large temperature range over which states in the excited
bands are unaccessible, yet all the ground band states are uniformly
occupied. The entropy value indicated by the dashed line in Figs.
\ref{FIG:ST1}-\ref{FIG:ST3} corresponds to the total number of $N_{p}$-particle
states in the ground band. Since the number of single particle energy
states in the ground band is equal to the number of lattice sites,
the total number of available $N_{p}$-particle states, which we define
as $\Omega_{0}$, is the number of distinct ways $N_{p}$ identical
bosons can be placed into $N_{s}$ states, i.e.
\begin{equation}
\Omega_{0}=\frac{(N_{p}+N_{s}-1)!}{N_{p}!(N_{s}-1)!}.
\label{eq:Omega0}
\end{equation}
The associated entropy $S_{0}=k_{B}\log\Omega_{0}$, which we shall
refer to as the plateau entropy, can be evaluated using Sterling's
approximation \begin{eqnarray}
S_{0} & \simeq & k_{B}\Big[N_{p}\log\left(\frac{N_{p}+N_{s}-1}{N_{p}}\right)\label{eq:Splateau}\\
 &  & +(N_{s}-1)\log\left(\frac{N_{p}+N_{s}-1}{N_{s}-1}\right)\Big].\nonumber \end{eqnarray}
As described earlier, this entropy value separates the heating and
cooling regions for the lattice.

\subsection{Scaling}

Here we give limiting results for the entropy-temperature curves.

\subsubsection{$k_{B}T\ll\epsilon_{{\rm BW}}.$}

When the temperature is small compared to the ground band width, only
energy states near the ground state of the lowest band are accessible
to the system. These states exhibit a quadratic dependence on the
magnitude of the quasimomentum and the entropy of this system is well
described by the free particle expression if the bare particle mass
is replaced by the effective mass i.e. $m\to m^{*}$. In general this
regime occurs only at very low temperatures, and except for extremely
low filling factors the system will be condensed in this region. In
this limit the expression for entropy corresponds to the expression
for a condensed gas of free particles with mass $m^{*}$ 

\begin{equation}
S=\frac{5}{2}N_{s}a^{3}\zeta(5/2)\left(\frac{m^{*}k_{B}T}{2\pi\hbar^{2}}\right)^{\frac{3}{2}},\label{eq:S_lowT}\end{equation}
where $\zeta(n)=\sum_{k=1}^{\infty}1/k^{n}$ is the Riemann Zeta function
and $N_{s}a^{3}$ is the volume of the system \cite{Pethick2002a}.
We note that in this regime the critical temperature for condensation
scales as $T_{c}\sim(m^{*})^{-3/2}$, so that $T/T_{c}$ is independent
of $m^{*}$ and hence the lattice depth.

\subsubsection{Tight-binding limit with $k_{B}T\ll\epsilon_{{\rm gap}}$ \label{sub:Tight-binding-limit-with}}

As discussed in Sec. \ref{sub:Entropy-Plateau}, when the temperature
is small compared to the energy gap only states within the ground
band are accessible to the system. In addition when the tight-binding
description is applicable for the initial and final states of an adiabatic
process, the initial and final properties are related by a scaling
transformation.

To illustrate we consider our initial system to be in equilibrium
with entropy $S_{i}<S_{0}$, in lattice of depth $V_{i}$ which we
take to be sufficiently deep enough for tight-binding expression (\ref{eq:ETB})
to be a good description of the ground band energy states. If an adiabatic
process is used to take the system to some final state at lattice
depth $V_{f}$ (also in the tight-binding regime) it is easily shown
that the macroscopic parameters of the initial and final states are
related as\begin{eqnarray}
E_{f} & = & \alpha E_{i},\label{eq:scaleE}\\
T_{f} & = & \alpha T_{i},\label{eq:scaleT}\\
\mu_{f} & = & \alpha\mu_{i},\label{eq:scalemu}\end{eqnarray}
where the scale factor \begin{equation}
\alpha=m_{i}^{*}/m_{f}^{*}\label{eq:alphadef}\end{equation}
 is the ratio of the effective masses, and the subscripts $i$ and
$f$ refer to quantities associated with the initial and final states
respectively. This type of scaling relation leaves the occupations
of the single particle levels (including the condensate occupation)
unchanged. This means that being adiabatic does not require redistribution
through collisions and may allow the lattice depth to be changed more
rapidly. However, adiabaticity with respect to the many-body wavefunction
will become a significant constraint as the lattice depth increases
and will become the limiting timescale.

\subsection{Condensation}

The temperature and entropy at which bosons condense generally changes
with lattice depth as indicated in Figs. \ref{FIG:ST1}-\ref{FIG:ST3}.
We note that for high filling factors the condensation points for
different lattice depths occur over a wide range of entropies, suggesting
that the degree of condensation will be greatly affected by adiabatic
lattice loading. 

For instance, consider the adiabatic process indicated by the dotted
line and labeled $C$ in Fig. \ref{FIG:ST3}. The system starts as
a Bose-condensed gas of free particles. However, as the lattice depth
increases the condensate fraction decreases until the system passes
through the transition point and becomes uncondensed. This process
is reversible, and is analogous to the experiments by Stamper-Kurn
\emph{et al.} \cite{Stamper1998a}, where a Bose gas was reversibly
condensed by changing the shape of the trapping potential.

We note that the free particle case with $n=1$ shown in Fig. \ref{FIG:ST2}
condenses at an entropy approximately equal to the plateau entropy
$S_{0}$, given by Eq. (\ref{eq:Splateau}). For an ideal gas of free
particles condensation occurs when the entropy of the system is $S_{{\rm cond}}=\frac{5}{2}k_{B}N_{p}\zeta(\frac{5}{2})/\zeta(\frac{3}{2}),$
(e.g. see \cite{Pethick2002a}), whereas the plateau entropy for a
lattice with unit filling is $S_{0}(n\!=\!1)=2k_{B}N_{p}\log(2)$,
(obtained by setting $(N_{s}-1)\to N_{s}$ and $N_{p}=N_{s}$ in Eq.
(\ref{eq:Splateau})). In comparing the values we find $S_{0}(n\!=\!1)/S_{{\rm cond}}\approx1.0798$,
in agreement with numerical observation. For fixed $N_{s}$ the filling
factor is proportional to $N_{p}$, thus the free particle condensation
entropy scales as $S_{{\rm cond}}\sim n$ whereas the entropy plateau
goes as $S_{0}\sim\log(n)$. So we conclude that for $n\gtrsim1$
condensation occurs (for free particles) above the plateau, whereas
for $n\lesssim1$ it occurs below the plateau.

\subsection{Interaction effects\label{sub:Interaction-effects}}

We now turn to a qualitative discussion of interaction effects on
the system. To do this we employ Bogoliubov's method for treating
an interacting Bose gas. In this approach all the atoms are assumed
to be in the zero quasi-momentum condensate, and by diagonalizing
an approximation to the many-body Hamiltonian a set of quasiparticle
levels are obtained (e.g. see \cite{Sorensen1998a,Burnett2002a}).
This approach should be a good approximation to the exact many-body
spectrum for moderate lattice depths (i.e. below the depth where the
Mott-transition occurs) and small thermal depletion. 

\begin{figure}
\includegraphics[%
  width=3.4in]{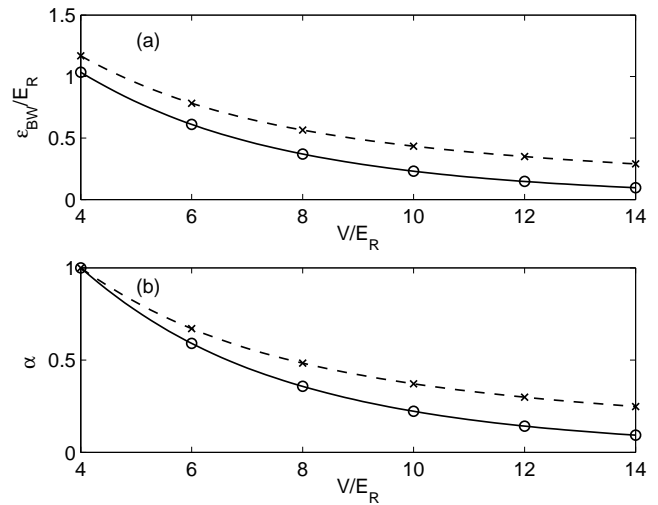}

\caption{\label{FIG:InterEffects} Bogoliubov description of interaction effects
on the (a) bandwidth and (b) scaling parameter $\alpha$ for various
lattice depths. Crosses: interacting result for $n=1$. Circles: non-interacting
result. Other parameters: $\lambda=850$nm and collisional interaction
strength taken for Rb$^{87}$.}
\end{figure}

The regimes of cooling associated with lattice loading observed in
Figs. \ref{FIG:ST1}-\ref{FIG:ST3} arose from the rapid compression
of the ground band width ($\epsilon_{{\rm BW}}$) that occurs with
increasing lattice depth. To assess the effects of interactions on
cooling we compare the width of the ground band calculated with Bogoliubov
theory to the non-interacting case for parameters relevant to current
experiments in Fig. \ref{FIG:InterEffects}(a). These results demonstrate
that interactions suppress the rate at which the bandwidth decreases
as the lattice is applied. We also note that the band gap is modified
by interactions, however the thermodynamic properties of the system
are not as sensitive to small changes in this and we will not consider
the effect of this. 

To quantify how the modified ground band excitation spectrum affects
thermodynamic properties of the system we consider an extension of
the arguments made in Sec. \ref{sub:Tight-binding-limit-with}. In
that section we discussed the scaling relationship that could be applied
to the thermodynamic quantities for a Bose gas in the tight-binding
regime when only states of the ground band were relevant. For the
interacting case we can apply the same argument if we ignore any ground
band spectrum reshaping with varying lattice depth, other than a scale
change given by the scaling parameter $\alpha=(\epsilon_{{\rm BW}})_{f}/(\epsilon_{{\rm BW}})_{i}$,
where $(\epsilon_{{\rm BW}})_{i}$ and $(\epsilon_{{\rm BW}})_{f}$
are the respective bandwidths of the initial and final configurations
(both assumed to be in the tightbinding limit). We note that this
definition of $\alpha$ is equivalent to Eq. (\ref{eq:alphadef})
for the non-interacting case. As was shown in Eq. (\ref{eq:scaleT}),
the ratio of final to initial temperatures is equal to $\alpha.$
In Fig. \ref{FIG:InterEffects}(b) we compare the values of $\alpha$
both with and without interactions for a system initially in a $V=4E_{R}$
deep lattice, for various final depths $V$ %
\footnote{We note that for the upper depth limit used in Figs. \ref{FIG:InterEffects}(a)-(b)
the system will likely be in the Mott-insulating state (for typical
experimental parameters and filling factors of order unity), in which case
the Bogoliubov approach will not  accurately
describe the excitation spectrum.%
}. These results quantify how interactions lead to a lesser degree
of cooling, and (in the limit of the Bogoliubov approach) these results
suggest that the temperature will level out to some finite non-zero
value at large lattice depths.

In the Mott-insulator regime (where Bogoliubov approach is invalid)
an energy gap proportional to the on-site interaction strength develops
between the ground state and the particle-hole excitations above it
(see \cite{Jaksch1998a}). This suggests that in general interactions
tend to increase ground band width, and will limit our ability to
cool the system with adiabatic passage.

\subsection{Adiabaticity\label{sub:Adiabaticity}}

Finally we note that interactions between particles are essential
for establishing equilibrium in the system, and understanding this
in detail will be necessary to determine the timescale for adiabatic
loading. In general this requirement is difficult to assess, and in
systems where there is an additional external potential it seems that
the adiabaticity requirements will likely be dominated by the process
of atom transport within the lattice to keep the chemical potential
uniform, though recent proposals have suggested ways of reducing this
problem \cite{Sklarzr2002a}. 

It seems reasonable that for sufficiently deep lattices the decreasing
tunneling rate will ultimately become the rate limiting timescale
for maintaining adiabaticity in adiabatic loading (e.g. see \cite{Band2002a}).
To estimate this timescale we consider a case relevant to $^{87}$Rb
experiments. For a lattice depth of a $V=10E_{R}$, where we have
taken $\lambda=850$nm, the tunneling time is $\sim10$ms. This timescale
is short compared to the loading time used in recent experiments with
this system \cite{Orzel2001a,Greiner2002a}, and suggests that smoothly
increasing the lattice to depths of $\sim10E_{R}$ over $>100$ms
should be very adiabatic. For depths larger that $V=10E_{R}$ the
tunneling time increases exponentially and the adiabatic condition
becomes more difficult to satisfy, however this is also the regime
where many-body effects will begin to dominate and a more complete
description will be needed to fully understand the adiabaticity requirements. 

\begin{figure}
\includegraphics[%
  width=3.5in]{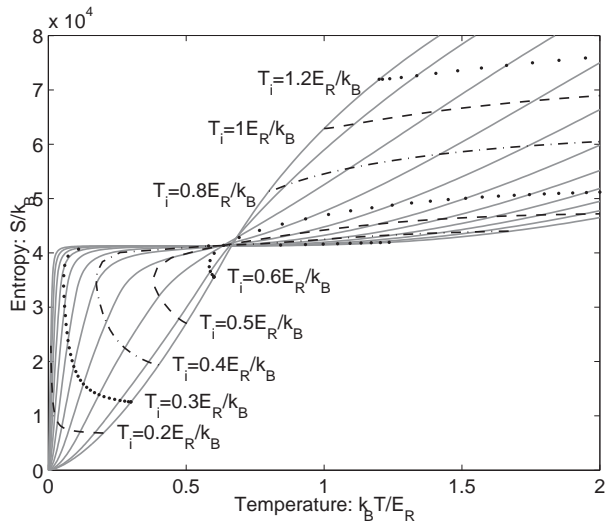}

\caption{\label{FIG:nonadiab} Fast lattice loading of a $N_{s}\approx3\times10^{4}$
site cubic lattice, with filling factor $n=1$. Broken/dotted/dashed
lines indicate fast loading curves (see text) and are labeled by the
temperature of the initial $V=0$ state. The lattice depth on these
curves can be determined from their intercept with the equilibrium
entropy versus temperature curves (gray solid lines), which are described
in Fig. \ref{FIG:ST2}.}
\end{figure}

It is useful to assess the degree to which non-adiabatic loading would
cause heating in the system. We consider lattice loading on a time
scale fast compared to the typical collision time between atoms, yet
slow enough to be quantum mechanically adiabatic with respect to the
single particle states. This latter requirement excludes changing
the lattice so fast that band excitations are induced, and it has
been shown that in practice this condition can be satisfied on very
short time scales \cite{Denschlag2002a}. We will refer to this type
of loading as fast lattice loading, to distinguish it from the fully
adiabatic loading we have been considering thus far.

To simulate the fast lattice loading we take the system to be initially
in equilibrium at temperature $T_{i}$ for zero lattice depth. For
the final lattice depth we fast load into, we map the initial single
particle distribution onto their equivalent states in the final lattice,
and calculate the total energy for this final non-equilibrium configuration.
This procedure assumes that there has been no collisional redistribution
to allow the system to adjust to the changing potential. To determine
the thermodynamic state the final distribution will relax to, we use
the energy of the non-equilibrium distribution as a constraint for
finding the equilibrium values of temperature and entropy. In general
the final state properties will depend on the initial temperature,
filling factor, and final depth of the lattice, and to illustrate
typical behavior we show a set of fast loading process curves that
indicate the final state equilibrium properties as a function of final
lattice depth in Fig. \ref{FIG:nonadiab} for unit filling and various
initial temperatures.

These curves show, as is expected from standard thermodynamic arguments,
that entropy always increases for non-adiabatic processes, i.e. all
loading curves in Fig. \ref{FIG:nonadiab} bend up wards with increasing
lattice depth. We also observe that eventually all curves predict
heating for large enough final lattice depth. However, for initial
temperatures sufficiently far below the critical temperature for cooling
$(T_{0}\sim0.7E_{R}/k_{B}$) a useful degree of temperature reduction
can be achieved with fast lattice loading up to certain maximum depth.
For example, the curve with initial temperature $T_{i}=0.5E_{R}/k_{B}$
in Fig. \ref{FIG:nonadiab} cools for final depths less than $V\sim5E_{R}$,
but for depths greater than this the temperature begins to increase
quite rapidly.

\section{2D Lattice with dipole confinement}

\begin{figure}
\includegraphics[%
  width=3.4in]{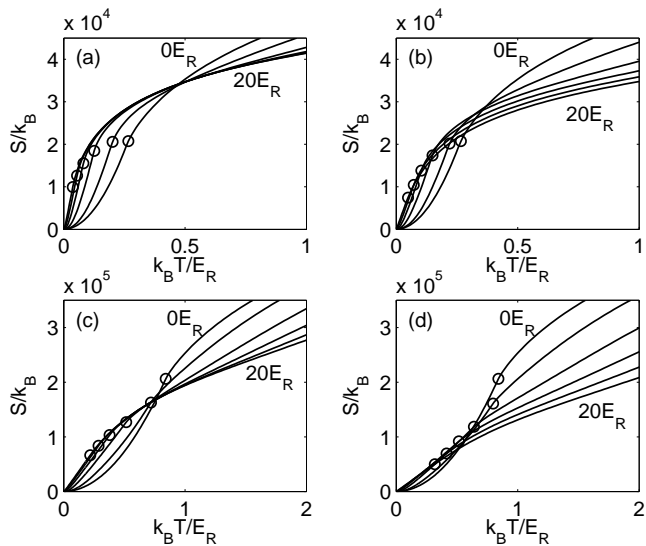}

\caption{\label{FIG:ST2D} Entropy versus temperature curves for a $N_{s}\approx10^{2}$
site 2D square lattice with harmonic trapping potential in the perpendicular
direction, for various lattice depths and strengths of harmonic confinement.
Results are given 2D site occupations of (a)-(b) $n=10$, and (c)-(d)
$n=100$. In each subplot the various curves correspond to lattices
depths of $V=0$ to $20E_{R}$ (with a spacing of $4E$ between each
curve). In (a) and (c) the harmonic confinement frequency is $f_{T}=0.01E_{R}/h$,
whereas in (b) and (d) $f_{T}=0.01E_{R}/h$ at $V=0$, and is taken
to increase by $2\times10^{-3}E_{R}$ for each subsequent lattice
depth, to $f_{T}=0.02E_{R}$ at $V=20E_{R}$. The condensation point
for the system is marked with a circle on each curve.}
\end{figure}

Here we consider the case of a 2D lattice in the $xy$-plane with
a harmonic trapping potential along the $z$ direction. We calculate
the eigenstates of the Schr\"odinger equation with potential energy

\[
V_{T}(\mathbf{r})=\frac{V}{2}[\cos(2kx)+\cos(2ky)]+\frac{1}{2}m(2\pi f_{T})^{2}z^{2},\]
where $f_{T}$ is the harmonic trap frequency. In Fig. \ref{FIG:ST2D}
we give the entropy versus temperature curves for a range of experimentally
relevant parameters. The results in Figs. \ref{FIG:ST2D} (a) and
(c) demonstrate the behavior for a range of lattice depths and two
different atom densities, but with constant harmonic trap frequency.
These curves exhibit qualitative similarities with the 3D lattice
curves shown in Figs. \ref{FIG:ST1}-\ref{FIG:ST3}, such as a crossing
region where the zero-depth curve changes from being a lower bound
to an upper bound of the other curves, and separates the regions where
adiabatic loading of the system will cool or heat the system respectively.
Similar to the 3D case, this behavior arises because of how the lattice
modifies the energy spectrum, i.e. the compression of low lying states
and upward shift of high lying states. However, the availability of
equally spaced harmonic oscillator states ensures that the density
of states does not have a gap (assuming $f_{T}$ is small compared
to $V$) and an entropy plateau does not appear. We note that the
condensation temperature decreases more rapidly both with temperature
and entropy compared to the 3D cases, making the 2D lattice a more
ideal system for observing reversible condensation.

In Figs. \ref{FIG:ST2D} (b) and (d) we consider equivalent systems
to those in Figs. \ref{FIG:ST2D} (a) and (c), except that we increase
the trap frequency with lattice depth to model the effects of additional
dipole confinement on the system. For the case of rubidium and taking
$\lambda\sim805$nm, these results correspond to a harmonic confinement
of $f_{T}\sim35$Hz at zero lattice depth, with the confinement increasing
in linear steps on successive curves up to a maximum value $f_{T}\sim70$Hz
at $V=20E_{R}$, typical of current experimental parameters. These
results demonstrate that the additional dipole confinement reduces
the size of the region over which cooling occurs, and reduces the
extend to which the system can be cooled. The more rapidly $f_{T}$
increases with lattice depth, the more pronounced this reduction will
be.

\section{Conclusion}

In this paper we have calculated the entropy-temperature curves for
bosons in a 3D optical lattice, and a 2D lattice with harmonic confinement
for various depths and filling factors. We have identified general
features of the thermodynamic properties relevant to lattice loading,
indicated regimes where adiabatically changing the lattice depth will
cause heating or cooling of the atomic sample, and have provided limiting
results for the behavior of the entropy curves. We have considered
the effect of lattice depth and filling factor on the Bose condensation
point and have examined the possibility of reversible condensation
through lattice loading. We have discussed the dominant effects of
interactions, and have shown that many of our predictions are robust
to non-adiabatic effects. Future extensions to this work will consider
in more detail the effects of both interactions and inhomogeneous
external potentials.

\section*{Acknowledgments}

The authors acknowledge useful discussions with S. L. Rolston, and
thank Charles Clark for useful comments. This work was supported by
the US Office of Naval Research, and the Advanced Research and Development
Activity.

\end{document}